\documentclass{jkas}

\def\year{2014} 
\def\month{March} 
\def\volume{00} 
\def\issue{0} 
\def\beginpage{1} 
\def\endpage{99} 
\def\received{---} 
\def\accepted{---} 

\date{Received \received ; accepted \accepted}

\title{
{\it WFIRST} Ultra-Precise Astrometry I: Kuiper Belt Objects
}

\author{Andrew Gould}

\affil{Department of Astronomy Ohio State University,
140 W.\ 18th Ave., Columbus, OH 43210, USA 
\email{gould@astronomy.ohio-state.edu }}

\def\corrauthor{
Andrew Gould
}

\def\runningauthor{
Andrew Gould
}

\def\runningtitle{
KBOs from WFIRST Astrometry
}

\def\keywords{
astrometry -- Kuiper belt -- gravitational microlensing 
}

\def\abstracttext{
I show that the {\it WFIRST} microlensing survey will enable
detection and precision orbit determination of Kuiper Belt Objects
(KBOs) down to $H_{\rm vega}=28.2$ over an effective area of 
$\sim 17\,\rm deg^2$.  Typical fractional period errors will be 
$\sim 1.5\%\times 10^{0.4(H-28.2)}$ with similar errors in other parameters
for roughly 5000 KBOs.
Binary companions to detected KBOs can be detected to even fainter limits,
$H_{\rm vega}=29$, corresponding to $R\sim 30.5$ and effective diameters
$D\sim 7\,$km.  For KBOs $H\sim 23$,
binary companions can be found with separations down to 10 mas.
This will provide an unprecedented probe
of orbital resonance and KBO mass measurements.
More than a thousand stellar occultations by KBOs can be combined to
determine the mean size as a function of KBO magnitude down to $H\sim 25$.
Current ground-based microlensing surveys can make a significant
start on finding and characterizing KBOs using existing and soon-to-be-acquired
data.
}

\newcommand{\bdv}[1]{\mbox{\boldmath$#1$}}

\def\au{{\rm AU}} 
\def\bx{{\bf x}} 
\def\bv{{\bf v}} 
\def\vega{{\rm vega}} 
 
\def\snr{{\rm SNR}} 
 
\def\kms{{\rm km}\,{\rm s}^{-1}}
\def\masyr{{\rm mas}\,{\rm yr}^{-1}}
\def\muas{{\mu\rm as}}

\def\ast{{\rm ast}}

\def\bnu{{\bdv\nu}}
\def\btheta{{\bdv\theta}}

\def\pix{{\rm pix}}
\def\hbn{{\hfil\break\noindent}}
\def\la{{<\atop \sim}}
\def\ga{{>\atop \sim}}
\def\apj{{ApJ}}
\def\aj{{AJ}}

\begin{document}
\jkashead 


\section{{Introduction}
\label{sec:intro}}

Kuiper Belt Objects (KBOs) provide an extraordinary probe of the
origin and history of the Solar System.  When Pluto was discovered
by Clyde Tombaugh \citep{slipher30,slipher30b}
and was then found to be in a 3:2 resonance with Neptune,
it was hardly guessed that it was only the largest of a vast class
of such objects.  Subsequent discovery of KBOs in 2:1 resonance,
in various kinematic and composition subclasses, of binary KBOs,
and of a break in the size distribution at $R\sim 26.5$ \citep{bernstein04}
have placed extremely detailed constraints on early Solar System
evolution, even leading to radical conjectures like the idea
that Uranus and Neptune originally formed much closer to the orbits of
Jupiter and Saturn \citep{nice}.  I show that the {\it WFIRST} microlensing
survey will, without any adjustment, yield a KBO survey that is
both substantially deeper and three orders of magnitude
wider and more precise than
existing deep KBO surveys based on {\it Hubble Space Telescope (HST)}
data.

\section{{KBOs in Microlensing Data}
\label{sec:kboulens}}

Microlensing surveys are conducted toward the densest star fields
on the sky and therefore would appear at first sight to be the
worst place to carry out a search for KBOs.  In particular, confusion
in the identification of moving objects is known to be an extremely
strong function of the density of the stellar background 
(e.g., \citealt{usnob}).

Surprisingly, microlensing fields are actually the best place to
conduct deep searches for moving objects, KBOs in particular.
First, contrary to naive expectation, there is no problem of confusion
at all.  Microlensing searches are conducted on a series of 
{\it difference images}, which are constructed by first forming a
deep (essentially noiseless) reference image from many individual
images, aligning each image geometrically to the reference image,
convolving the reference image to the point spread function (PSF) 
of each image, aligning each image photometrically to the reference
image, and finally subtracting each image from the reference
image \citep{alard98}.
The result is a virtually flat image with
essentially all stars removed and only photon and detector noise
left.  The only exceptions are stars (or other objects) that have
changed relative to the reference image, either changed their
brightness or changed their position, and a very small number of
artifacts due to difficulties in subtracting extremely bright and/or
saturated stars.  Hence, microlensing fields are not crowded at all:
they are essentially blank, much more so than high-latitude fields.

Second, the noise properties of every pixel (as a function of both
position on the sky and position within the detector array) are
understood essentially perfectly.  This is because there are hundreds,
thousands, or even tens of thousands of images of the same field.
In the case of ground-based surveys, these encompass the full range
of observing conditions, while for space-based surveys, the images
are taken under constant conditions (or rather conditions that vary
much less than is relevant to faint objects).  

Third, microlensing surveys carry out hundreds or thousands of
observations during a single season (during which KBOs 
remain mostly within the microlensing fields).  While each
individual exposure is not particularly deep, the fact that the images
are essentially blank means it is straightforward to combine many
observations to detect faint KBOs.  In particular, while the KBO
will occasionally ``land'' at the position of a relatively
bright star, the precise knowledge of high noise at this location 
(derived from
photometric deviations on hundreds of other images -- see previous
paragraph) will automatically ``suppress'' this observation within
the ensemble.  That is, KBO searches will automatically be dominated
by the highest signal-to-noise ratio (SNR) observations.

Fourth, because the Galactic Center is only about $6^\circ$ south
of the ecliptic, microlensing surveys automatically probe regions
of the sky on and near the ecliptic.

Finally, the high density of stellar background is actually an advantage
because it gives rise to occultations that can be used to measure
the size distribution.

To date, no one has yet suggested, let alone explored, the possibility
of KBO searches in microlensing fields. Here, I focus on the potential
of future space based surveys, but the same principle applies
to ground-based data.  In particular, the Optical Gravitational Lens
Experiment (OGLE) survey already has
more than 4 years of data covering about $5\,{\rm deg}^2$ with a cadence of
roughly 30 observations per night during peak microlensing season,
with typical seeing slightly more that $1^{\prime\prime}$,
while the new Korean Microlensing Telescope Network
(KMTNet) survey is expected to cover $16\,{\rm deg}^2$ with
a cadence of about 120 observations per night with only slightly
worse seeing (counting only Chile and South Africa telescopes).

While no KBO searches have been conducted using microlensing data,
OGLE did search for KBOs in the $2500\,\rm deg^2$ of the southern
Galactic plane \citep{ogleplane}.  Because there were only three
exposures, each 180 s
(for the overwhelming majority of fields), the survey reached only 
$R=21.6$.  Highly illustrative is their ``rediscovery''
of Pluto at roughly $(l,b)=(13,-2)$ in a dense stellar field (their Figure~3),
which becomes essentially blank (except for Pluto) in the difference
image (their Figure~4).  


\section{{Astrometry From A Microlensing Survey}
\label{sec:ulens}}

A space-based microlensing survey will almost automatically return 
high-quality astrometric data.  I discuss this potential within the
framework of the proposed {\it WFIRST} survey, but the same principles
could be applied to any mission of this type.  I adopt the following
parameters for the microlensing component of the proposed {\it WFIRST} 
mission.  
\hbn 1) 10 contiguous, $0.28\,\rm deg^2$ Galactic bulge fields
\hbn 2) six $\Delta t=72\,$day continuous ``campaigns'' , 
each centered at quadrature (March or September)
\hbn 3) 15 min cycle time consisting of ten 52s exposures
\hbn 4) 90\% of exposures in broad $H$ band, 10\% in a narrower band, e.g., $Y$
\hbn 5) 2.4m telescope, $\theta_{\rm pix}=110\,$mas pixels,
\hbn 6) one photon per second  at $H_\vega=26.1$
\hbn 7) $B=341$ total ``counts'' per pixel in read noise, zodiacal light and 
dark current per 52s integration

The diffraction limited point spread function
(PSF) has FWHM$=275\,$mas, implying an equivalent
Gaussian width $\sigma_{\rm psf} = {\rm FWHM}/2.35=75\,$mas.  Due to the
slightly undersampled PSF, I assume a total background light of $9B$, i.e.,
about 1.5 times larger than the $4\pi(\sigma_{\rm psf}/p)^2B$ appropriate
to the oversampled limit.  This leads to an equivalent ``sky'' of
$H_{\rm sky} = 26.1 - 2.5*\log(9B/52) = 21.7$.  Since this paper will
concern only sources well below this limit, I restrict further consideration
to ``below sky'' sources.  For these, the signal-to-noise ratio (SNR) is
given by
\begin{equation}
\snr = 10^{0.4(H_{\rm zero} - H)};\qquad H_{\rm zero} = 26.1
\label{eqn:snr}
\end{equation}

The 15 min cadence implies a total of 6908 observations during a 72 day
campaign.  Of these 10\% are not in the wide band and so contribute
much less astrometric information.  These will simply be ignored
(though they will provide important color information).
I further assume that 10\% of the remainder land on sources that
are significantly above sky and therefore have substantial additional
noise in the difference images.  This leaves 
\begin{equation}
N_{\rm cam} = 5600
\label{eqn:ncam}
\end{equation}
images, distributed roughly uniformly 
in time over each campaign.  I assume
that each of these has astrometric precision
\begin{equation}
\sigma_{\rm ast} = \sqrt{2}{\sigma_{\rm psf}\over \snr}={106\,{\rm mas}\over \snr}.
\label{eqn:sigmaast}
\end{equation}

The images will be dithered and the KBOs
will move relative to the stellar background, so the
astrometric and photometric errors of the $N_{\rm cam} = 5600$ images will
be essentially uncorrelated.  Because of the very large number
of images, the central limit theorem therefore implies that Gaussian
statistics {\it strictly} apply.

\section{{World Coordinate System}
\label{sec:wcs}}


In order to fit KBO orbits over the $\sim 1^\circ$ trajectories that they
traverse during a 72-day campaign, the {\it WFIRST} astrometric frame
must be calibrated to at least the precision of the KBO measurements.
Since GAIA will measure positions, parallaxes, and proper motions
for several million stars in the {\it WFIRST} microlensing fields, it 
appears at first sight that this issue will be taken care of
``automatically''.  While this ultimately proves to be the case,
there are some subtleties, which I now address.

The central difficulty is that {\it WFIRST} will ``saturate'' 
(see below) at $H_\vega\sim 14$,
while the majority of GAIA stars will be bulge sub-giants and
giants with $(V-H)_0 > 1.5$
that lie behind most of the dust with $E(V-H)\sim 3.5$.  For such
stars to avoid saturation in {\it WFIRST}, they must be $G>19$ in the GAIA
band (which is near $V$).  While GAIA will observe stars to significantly
fainter magnitudes over most of the sky, it will have a substantially 
brighter limit in the bulge due to data-rate limitations.  Hence,
it is not immediately obvious that the {\it WFIRST} frame can be
tied to the GAIA frame.

The best frame-tying stars are foreground G dwarfs, with $(V-H)_0\sim 1.5$
and $M_V\sim 5$.  I assume $A_V=0.7\,\rm mag\,kpc^{-1}$ and $A_V/A_H=6$.
Such stars reach the {\it WFIRST} saturation limit at $D\sim 1.2\,$kpc
at which $H\sim 14$ and $V\sim 16.2$, while at $D\sim 1.8\,$kpc, the 
corresponding numbers are $H\sim 15$ and $V\sim 17.5$.  These stars
will have GAIA $G$ band magnitudes $16.1<G<17.2$ and hence
GAIA position,
parallax, and proper motion 
precisions\footnote{http://www.cosmos.esa.int/web/gaia/science-performance} 
ranging
from $(32,43,22\,{\rm yr}^{-1})\muas$ to $(54,73,39\,{\rm yr}^{-1})\muas$.

To set up a reference frame, all we are really concerned with is
how well the positions of these stars will be known roughly $T=10\,$yr
after the midpoint of the GAIA mission when {\it WFIRST} is launched.
({\it WFIRST} proper motions of these overlap stars
will be measured far better than those
of GAIA, so we are concerned with overlap at the beginning of the 
{\it WFIRST} mission, not its midpoint.)\ \ Hence, the positions will
be known to a precision of 200--400$\,\muas$.

The surface density of such stars on the sky is
$N=n(D_{\rm max}^3-D_{\rm min}^3)/3 = 4100\,{\rm deg}^{-2}$, 
where $n=0.01\,{\rm pc}^{-3}$ is the number density of G dwarfs at the
mean distance.

These overlap stars must be used for two types of alignment.  First,
the 10 {\it WFIRST} fields must be aligned with each other.  Second,
the ``pixel frame'' of the camera must be calibrated to an externally
determined angular scale.  Since each frame has $0.28\times 4100\sim 1150$
stars, it can be aligned to the global system to 
$300\,\muas/\sqrt{1150}\sim 9\,\muas$.  For purposes of internal alignment,
all the overlap stars observed in the 10 fields can be used, implying
a density of $\sim 41000\,{\rm deg}^{-2}= (18^{\prime\prime})^{-2}$.

At what point do each of these two calibration issues become the limiting
factor?  Each KBO spends roughly 25 days (or 1800 observations) inside
a field, which implies that a $9\,\muas$ calibration error becomes an
issue when the individual astrometric error falls below 
$9\,\muas\sqrt{1800}= 380\,\muas$.  This occurs at $\snr=280$, i.e.,
$H=20$, far brighter than any KBO of practical interest.  
Similarly, the KBO remains in the neighborhood of a calibration
star for $18^{\prime\prime}$, corresponding to about 25 observations,
so calibration would become an issue at $\snr=60$ corresponding to
$H=21.7$, still much too bright to be of concern.

I note that {\it WFIRST} microlensing fields will be dithered, so
that 40,000 images of each field will permit high-precision calibration
of the pixel scale on the $18^{\prime\prime}$ scale of the typical
separations between calibrating stars.

Finally, I return to the question of the {\it WFIRST} saturation limit.
The arrays go non-linear at about half the full well of 100,000 
photo-electrons.  In the broad $H$ filter no more than 30\% of
photons land in a single pixel.  The pixels are read out every 2.6 s.
These numbers lead to a conservative saturation estimate of
$H = 26.1 - 2.5\log(50000/(0.3*2.6) = 14.1$.  The question of what
is ``saturated'' depends strongly on the application.  If one is
interested in precise individual photometric measurements, then probably several
reads are required, not just the single read in the above calculation.
However, if one is interested solely in determining the astrometric
center of the PSF to, say, 1\% of a pixel, it is doubtful whether it is
actually necessary that the central pixel be unsaturated, let alone
in the linear regime, since the position can be centroided from other
pixels.  Further, one could restrict attention to the subset of observations
that land near the corner of four pixels.  Thus, it is quite plausible
that the saturation limit for astrometry could go 1 mag or more brighter
than $H=14$.  However, the above calculation shows that for present
purposes, such a detailed investigation is unnecessary.

\section{{Precision of Orbit Determination}
\label{sec:orbit}}

Assuming that a KBO is identified in the data, how well can its orbit
be measured?  To answer this question, we must first ask how long
it will stay within the field of view.  The first point is that
at semi-major $a\sim 40\,\au$, and hence period 
$P=(a/\au)^{3/2}{\rm yr}\sim 250\,$yr, a KBO will move about 1.4 deg
from one year to the next, and hence will typically not return to
the microlensing fields the next year.  On the other hand, since
the fields are observed at quadrature when Earth is accelerating
transverse to the line of sight at $A_\oplus=4\pi^2\au\,{\rm yr}^{-2}\sim
0.5\,\kms\,{\rm day}^{-1}$ while the KBO will be moving of order
$v_\perp\sim 5\,\kms$, the net projected relative motion of 
Earth against the KBO will be
\begin{equation}
\Delta x\sim \sin\biggl({2\pi(\Delta t/2+v_\perp/\au)\over{\rm yr}}\biggr)\au
\sim 0.71\,\au,
\label{eqn:deltap}
\end{equation}
which corresponds to 1.0 deg.  Hence, of order 60\% of the KBOs will
remain in the field for the duration of the campaign.  I will initially
restrict attention to this subgroup and reserve discussion regarding the
remainder to Section~\ref{sec:edge}.

To estimate the precision of the orbital parameters, I approximate
Earth as being in a circular orbit and approximate the KBO physical motion
during the period of observation as uniform.  The deviations from both
assumptions are slight and, what is more important, deterministic with
respect to the adopted parameters.  For example, the acceleration of the KBO
is given directly by its distance and position on the sky.  Thus, making 
these assumptions only slightly changes, but vastly simplifies the
``trial functions'' and hence renders tractable the error estimates while
not significantly impacting these estimates.  I follow \citet{gould13}
in making the initial estimates in Cartesian phase-space coordinates
(instantaneous positions and velocities) rather than the traditional
orbital invariants.  Of course, actual fits to data will use
Kepler invariants, but the Cartesian approach is more closely matched
to short timescale observations and therefore facilitates both
deeper understanding and simpler results.  The implications for
Kepler invariants are then easily derived.

The KBO then has motion $\bx(t) = \bx_0 + \bv t$, described by
six parameters $(\bx_0,\bv)=(r_0,\bx_\perp,v_r,\bv_\perp)$.  
I adopt the midpoint of the campaign as the origin of time.
For convenience, these six parameters
can be re-expressed as $(\Pi,\btheta_0,\bnu_\perp,\nu_r)$,
\begin{equation}
\btheta_0 \equiv {\bx\over r_0};
\quad
\bnu_\perp\equiv {\bv_\perp\over \Omega r_0};
\quad
\Pi\equiv {\au\over r_0};
\quad
\nu_r\equiv {v_r \over \Omega r_0}.
\label{eqn:parms}
\end{equation}
where $\Omega=2\pi\,\rm yr^{-1}$ is Earth's orbital frequency.
The {\it WFIRST} fields
are only a few degrees from the ecliptic, and for simplicity I assume
that the KBO is directly on the ecliptic.

The first four of these parameters in Equation~(\ref{eqn:parms})
are essentially direct observables, i.e.,
the position and instantaneous (normalized) proper motion of 
the KBO at the zero point of time.
The last two pose the main challenge.  Since these are derived entirely
from the motion of the KBO within the ecliptic plane, I restrict attention to
these two dimensions (radial and 1-D transverse).  Then the equation
for the angular position $\theta(t)$ is
\begin{eqnarray}
\theta(t) &=& {x_0 + v_\perp t - \au (\cos\Omega t -1)\over
r + v_r t -\au\sin\Omega t} \nonumber \\
&=& {\theta_0 + \nu_\perp \Omega t + 2\Pi\sin^2(\Omega t/2)\over
1 + \nu_r \Omega t - \Pi\sin\Omega t}.
\label{eqn:thetat}
\end{eqnarray}
Since there are four parameters to be determined, we should
expand to third order in time
\begin{equation}
\theta(t) = \sum_{i=0}^3 a_i (\Omega t)^i
\label{eqn:thetat2}
\end{equation}
where
\begin{eqnarray}
a_0 &=&  \theta_0
\quad
a_1 = \nu_\perp + \theta_0 Z
\quad
a_2 = 0.5\Pi + \nu_\perp Z + \theta_0 Z^2 \nonumber\\
a_3 &=& -0.5\Pi \nu_r + 0.5\Pi^2 + \Pi^2\nu_\perp -2\Pi\nu_\perp\nu_r 
\nonumber \\ 
&\quad& + \nu_r^2\nu_\perp + \theta_0(Z^3 - \Pi/6)
\label{eqn:adefs}
\end{eqnarray}
and $Z\equiv \Pi -\nu_r$.  To a good approximation the four
coefficients are well represented by their leading terms
\begin{equation}
a_0 =  \theta_0
\quad
a_1 \rightarrow \nu_\perp
\quad
a_2 \rightarrow 0.5\Pi
\quad
a_3 \rightarrow -0.5\Pi\nu_r .
\label{eqn:adefs2}
\end{equation}
For a uniform set of $N$
observations over time $\Delta t$, the covariance matrix for
these four coefficients is given by (e.g., \citealt{gould04}),
\begin{eqnarray}
c_{ij} &=& {\sigma_{\rm ast}^2\over N_{\rm cam}}(\Omega\Delta t)^{-(i+j)}\tilde c_{ij}
\nonumber \\
\qquad
\tilde c_{ij} &=& \left(\matrix{
9/4 & 0 & -15 & 0 \cr
0 & 75 & 0 & -420 \cr
-15 & 0 & 180 & 0 \cr
0  & -420 & 0 & 2800}\right).
\label{eqn:cvals}
\end{eqnarray}
Therefore, the errors in the three angular variables of
interest (i.e., excluding $\theta_0$) are
\begin{eqnarray}
{\sigma(\nu_\perp)\over\Pi^{3/2}} &=&\sqrt{75\over N}
(\Omega\Delta t)^{-1}\Pi^{-3/2}\sigma_\ast,
\nonumber \\
{\sigma(\Pi)\over\Pi} &=&\sqrt{720\over N}
(\Omega\Delta t)^{-2}\Pi^{-1}\sigma_\ast,
\nonumber \\
{\sigma(\nu_r)\over\Pi^{3/2}} &=&\sqrt{11200\over N}
(\Omega\Delta t)^{-3}\Pi^{-5/2}\sigma_\ast,
\label{eqn:evals1}
\end{eqnarray}
where in each case I have normalized to a relevant physical scale.
For typical parameters ($\Omega\Delta t\sim 1.25$, $\Pi\sim 1/40$),
the final expression in Equation~(\ref{eqn:evals1}) is larger than
either of the others by a factor $>300$.
This is the justification for ignoring the
correlations between these levels embedded in Equation~(\ref{eqn:adefs})
and simply using Equation~(\ref{eqn:adefs2}): the errors in $\nu_r$
completely dominate.  Translating to physical variables, we obtain
\begin{eqnarray}
& &{\sigma(v_\perp)\over v_\oplus\Pi^{1/2}} =\sqrt{75\over N}
(\Omega\Delta t)^{-1}\Pi^{-3/2}\sigma_\ast
\nonumber \\
&\rightarrow&
{1.2\times 10^{-5}\over \snr}
\biggl({N\over 5600}\biggr)^{-1/2}
\biggl({\Delta t\over 72\,\rm d}\biggr)^{-1}
\biggl({r\over 40\,\au}\biggr)^{3/2},
\nonumber \\
& &{\sigma(r)\over r} =\sqrt{720\over N}
(\Omega\Delta t)^{-2}\Pi^{-1}\sigma_\ast
\nonumber \\
&\rightarrow&
{4.8\times 10^{-6}\over \snr}
\biggl({N\over 5600}\biggr)^{-1/2}
\biggl({\Delta t\over 72\,\rm d}\biggr)^{-2}
\biggl({r\over 40\,\au}\biggr),
\nonumber \\
& &{\sigma(v_r)\over v_\oplus\Pi^{1/2}} =\sqrt{11200\over N}
(\Omega\Delta t)^{-3}\Pi^{-5/2}\sigma_\ast
\nonumber \\
&\rightarrow&
{3.8\times 10^{-3}\over \snr}
\biggl({N\over 5600}\biggr)^{-1/2}
\biggl({\Delta t\over 72\,\rm d}\biggr)^{-3}
\biggl({r\over 40\,\au}\biggr)^{5/2}.
\label{eqn:evals2}
\end{eqnarray}
These results imply that the orbit-parameter error ellipsoid is
essentially a 1-dimensional structure.  In Cartesian space, this
one dimension is associated with a single parameter: $v_r$.  After
transforming to Kepler coordinates, all parameters inherit this
error in $v_r$ but in a highly correlated way.  For example, for
roughly circular orbits, the period error $\sigma(P)$ is
\begin{equation}
{\sigma(P)\over P} = {3\over 2}\,{\sigma(a)\over a} \simeq
{3\over 2}\,{\sigma({\rm KE})\over {\rm KE}} \simeq
{3v_r\over v_\oplus\Pi^{1/2}}{\sigma(v_r)\over v_\oplus\Pi^{1/2}},
\label{eqn:pererr}
\end{equation}
where KE is the kinetic energy.
Hence, for typical values $v_r\sim 0.2 v_\oplus\Pi^{1/2}$, the fractional
period error is somewhat smaller than the last expression in
Equation~(\ref{eqn:evals2}).
In the next section, I will show that the theoretical limit
for finding KBOs is near $\snr\sim 1/7$.  Thus, even at this limit,
the period precision is of the order of 1.5\%.  At the break in the
KBO luminosity function, $R\sim 26.5$, 
so roughly\footnote{When making comparisons to optical measurements,
I adopt $R-H=1.4$, which is about 0.3 mag redder than the Sun.  
\citet{kbocolor} (Figures 2 and 3) show that there is a clustering of
KBOs near $(V-I,J-H)\sim (0.95,0.35)$, which when compared to
the solar values $\sim(0.68,0.32)$, indicates redder optical colors
but similar infrared colors to the Sun.  However, their Figure 2 also
shows that the overall optical color distribution is asymmetric,
with a second peak in the distribution, roughly 0.35 to the red
in $(V-I)$.  My adopted value $R-H=1.4$, which is meant only to
give an indication of more commonly used optical magnitudes, is
representative of these populations.  Note, however, at least some KBOs have
optical-infrared colors very similar to the Sun.  For example
Orcus has $V-J=1.08\pm 0.04$, compared to $(V-J)_\odot = 1.08$.} 
$H\sim 25.1$ or $\snr\sim 2.5$, the period error is 
$\sigma(P)/P\sim 0.09\%$.

\section{{Finding KBOs in the Data}
\label{sec:find}}

{\it WFIRST} may well be in geosynchronous orbit.  In principle,
this would add information to the measurements of $r$ and $v_r$.
However, as I show below, this added information plays an insignificant
role except in the margins of parameter space and was therefore
ignored in Section~\ref{sec:orbit}.  A geosynchronous
orbit would also somewhat complicate the search for KBOs in the
data and so cannot be completely ignored in the present section.
Nevertheless, I begin by ignoring it, partly to cover the
case of non-geosynchronous (e.g., L2) orbit and partly to be able
to show explicitly, further below, that the complications induced
by  geosynchronous orbit are not in fact significant.

The {\it WFIRST} field has $N_\pix =2.8\,\rm deg^2/(110\, mas)^2=3.0\times 10^9$
pixels.  Thus, for KBOs with 
\begin{equation}
\snr\ga \sqrt{2\ln{N_\pix\over\snr}} = 6.3,
\label{eqn:snrpix}
\end{equation}
it is possible to comfortably identify KBO candidates without
fear of massive contamination by noise spikes using a simple
two-dimensional (2-D) search over the image.  At this limit,
there may be some contamination, but this could easily be vetted
by examining successive images.  There are two points to make about
Equation~(\ref{eqn:snrpix}).  First, it assumes Gaussian statistics.
This may not be valid for the case of a 2-D search.  However, 
Equation~(\ref{eqn:snrpix}) primarily serves as an entry point
to the much larger (4-D, 5-D, and 6-D) searches that I describe 
below, for which Gaussian statistics are valid.  I therefore
ignore this complication.  Second, the $\snr$ appears on both sides
of the equation, meaning that the equation must be solved self-consistently.
This poses no actual difficulties, since it appears inside a rather large
logarithm factor on the rhs, but does call for an explicit remark.

Next, I consider 4-D searches over position (2-D, as above) and
proper motion (2-D).  I consider a search in a ``proper motion''
circle $\mu_0 = 12^{\prime\prime}\,\rm day^{-1}$ (relative to a KBO
in circular motion at $r=40\,\au$.
By conducting a 4-D search, I am implicitly
assuming that the other two Cartesian phase-space coordinates
($\Pi$ and $\nu_r$) are ``not important''.  Explicitly, this
assumption means that the parallax differences among the KBOs being
searched lead to angular displacements of $<1\,$pixel during the
time of the search.  For definiteness, I assume that KBOs of interest
have $\Pi<\Pi_0 = 0.03$.  The acceleration of Earth then leads to
a differential pixel displacement of 
\begin{equation}
\Delta\theta_\Pi = {1\over 2}(\Omega t)^2\Pi_0
= 8\theta_\pix \biggl({\Delta t\over \rm day}\biggr)^2
\label{eqn:pardtheta}
\end{equation}
due to parallax.  Thus, a 4-D search
is restricted to $\Delta t<9\,$hr, and therefore requires a
search radius of $\mu_0\Delta t$ and hence a total number 
\begin{equation}
N_\mu=\pi\biggl({\mu_0\Delta t\over\theta_\pix}\biggr)^2 = 
5300\biggl({\Delta t\over 9\,\rm hr}\biggr)^2
\label{eqn:nmu}
\end{equation}
of searches at each of $N_\pix=3\times 10^9$ pixels, for a total of
$N_{\rm try}=N_\pix N_\mu = 1.6\times 10^{13}$ searches.  In nine hours, there are
approximately $N_{\rm im}=30$ images.  Hence this yields a SNR 
threshold of
\begin{equation}
\snr\ga N_{\rm im}^{-1/2}\sqrt{2\ln{N_{\rm try}\over\snr}-\ln N_{\rm im}} = 1.38.
\label{eqn:snrtry}
\end{equation}

To dig to lower SNR, one must probe over longer durations, which requires
5-D or 6-D searches.  To find the boundary, I adopt a radial-velocity
search range $\Delta \nu_r = 1/300$ corresponding to 
$\pm 2\,\kms$ at $r=40\,\au$.  Thus the radial velocity becomes important
at $\Delta t\sim \theta_\pix/(\Delta v_r\Pi\Omega)\sim 9\,$hr, which is
nearly identical to the onset of extra searches due to parallax.
Combining all factors 
($(\Delta t)^2$ for proper motions, 
$(\Delta t)^2$ for parallax, and
$(\Delta t)^1$ for radial velocity), implies a search total of
\begin{equation}
N_{\rm try} = 7\times 10^{14}\biggl({\Delta t\over \rm day}\biggr)^5,
\label{eqn:search}
\end{equation}
implying a maximum search total for $\Delta t=72\,$day of 
$N_{\rm try} = 1.4\times 10^{24}$.  Applying Equation~(\ref{eqn:snrtry}),
with $N_{\rm im}=5600$, yields a threshold $\snr\ga 0.15$, and
so a theoretical limit of $H_\vega=28.2$, or 
roughly
$R\sim 29.6$.
Recall from Equation~(\ref{eqn:pererr}) that even such extreme
below-sky KBOs would have orbital parameter errors of order 1.5\%.

However, reaching this theoretical limit will be no picnic.  One
could convolve all the images with the PSF (e.g., \citealt{shao14,gould96}), 
so that each search
would require only $\sim 10 N_{\rm im}$ floating point operations,
and thus a total of $\sim 8\times 10^{28}$ operations.  
This should be compared to the $\sim 10^{12}$ floating point operations
per second (FLOPS) of a current graphics processing unit (GPU).
One might imagine assigning $10^4$ GPUs to this task for a year,
but this would only enable $3\times 10^{23}$ operations, which
is a factor $q\sim 2.5\times 10^5$ short of what appears to be
needed.

I will argue immediately below that this computational shortfall
can probably be bridged by a combination of several factors.
However, it is useful to ask how an arbitrary shortfall $q$ would
impact the depth of the survey.  From Equation (\ref{eqn:search}) 
the number of trials scales $N_{\rm try}\propto (\Delta t)^5$, while
the number of computations for each trial is linear in $\Delta t$.
Hence, a shortfall factor $q$ can be compensated by reducing 
$\Delta t$ by a factor $q^{1/6}=8(q/2.5\times 10^5)^{1/6}$.
Naively, this leads to an increase in the SNR limit by
a factor $q^{1/12}=2.8(q/2.5\times 10^5)^{1/12}$.  In fact, taking account
of the impact on the logarithmic factor due to the smaller number
of trials in Equation~(\ref{eqn:snrtry}), the actual degradation
is a factor 2.6, i.e., a limit $\snr=0.38$, corresponding to
$H<27.1$.  Note that while the KBO is found using a restricted
subset of the data, the full data set can still be used to
estimate orbital parameters, so that the estimates of precision given in 
Section~\ref{sec:orbit} remain valid.

For the survey to reach its maximum potential (i.e., 1 mag deeper than
the above limit) requires either greater computing power or better
search algorithms.  Since
the data will not be available for at least a decade, we should
fold in a ``Moore's Law'' factor of 30 (assuming doubling time of
2 years), but this is still only $\sim 10^{25}$ operations.  Since
the number of operations scales $\propto (\Delta t)^6$, this seems
to permit analysis of data intervals $\Delta t\la 16\,$days, so
only reaching $\snr\ga 0.32$ and limiting magnitude $H_\vega=27.4$.

In fact, it should be possible to go deeper using search techniques
that are more clever than brute force trials.  For example, one could
begin by restricting the search to $\Delta t= 16\,$days as above,
but initially cull out trajectories with $\Delta\chi^2>28$.  This
would capture exactly half of all ultimately recoverable KBOs
(i.e., those with $\snr>0.15$), since these would have 
$\langle \Delta\chi^2\rangle= 28$, while at the same time suffering
noise-spike contamination of ``only'' 
$\sim {\rm exp}(-28/2)/\sqrt{28}\sim 10^{-7}$.  Now, of course,
this would still result in $\sim 10^{14}$ noise spikes, but
these could be vetted fairly efficiently, as follows.  From 
Equation~(\ref{eqn:evals1}),
$\sigma(\Pi) = 4\times 10^{-5}$ and 
$\sigma(\mu)=11\,\rm mas\,day^{-1}$.
In fact, it is easily shown that the proper-motion error in the
direction perpendicular to the ecliptic is smaller by $\sqrt{12/75}=0.4$.
Therefore, even allowing for a $3\,\sigma$ range for these two quantities,
the total number of searches required for each such  ``preliminary candidate''
is only $\sim 10^7$.  If the procedure were repeated on 4 independent
subsamples, it would recover $1-2^{-4}\sim 94\%$ of all with $\snr>0.15$.
That is, almost full recovery with $\sim 8\times 10^{25}$ FLOPS rather than
$\sim 8\times 10^{28}$ required for a brute-force search.  That is,
this algorithmic improvement, by itself, pushes down the magnitude
limit by $\Delta H\sim (2.5/12)\log(1000)\sim 0.6\,$mag relative
to the brute force search.

One could imagine yet more clever ideas.  It is premature to work these
out in detail because the real algorithms would have to take account
of not only operation speed but also memory access for processors
that have not even been designed.  The point is that it is not unrealistic
to think that the theoretical limit of $\snr>0.15$ can be reached, or
at least approached within a few tens of percent.  To reiterate what
was said above, this limit corresponds to $H_\vega<28.2$ or roughly $R<29.6$,
with period errors $\sigma(P)/P\la 1.5\%$.  And even if these improvements
were not made, with present technology and a brute-force search one can still
reach $H_\vega<27.1$ or roughly $R<28.5$, with period errors 
$\sigma(P)/P\la 0.6\%$.  

\subsection{{Finding KBOs in Unexpected Orbits}
\label{sec:unexpected}}

In the above treatment, the emphasis was on finding extremely faint
KBOs, with SNR substantially 
below unity in individual exposures.  To this end, it
was necessary to restrict the search space to orbital parameters that
are in some ``expected'' range, partly to limit the frequency of noise
spikes, but mainly to make the search computationally tractable.  However,
it will also be possible to relax essentially all limits on orbital parameters
down to $H\sim 24.1$, which is at the detection threshold $\snr \sim 6.3$
required to distinguish a KBO from noise spikes in a single exposures.
See Equation~(\ref{eqn:snrpix}).  Now, as emphasized above,
Gaussian statistics do not apply to single images, but only to
ensembles.  However, the first point is that such non-Gaussian noise
spikes could easily be vetted by comparing to neighboring images.  Moreover,
it would be straightforward to conduct searches over 1 day intervals,
allowing for all parallaxes $\Pi<0.05$ and proper motions corresponding
to all bound orbits.  With about 90 exposures, such a search would
reach $\snr\sim 1$ corresponding to $H\sim 26$, i.e., about 1 mag below
the break.  If this search found significant numbers of objects in
unexpected (e.g., retrograde) orbits, then further searches could
be fine-tuned to find fainter KBOs in similar orbits.

\section{{Binary Companions and Mass Measurements}
\label{sec:mass}}


Regardless of the exact limit achievable for an ab initio search
for KBOs, it is possible to reliably identify binary companions to all 
those that are found down to $H_\vega\sim 29$, 
significantly fainter than the theoretical limit for isolated KBOs.
Detection of such binary companions will lead to mass estimates
and mass measurements of the parent KBO.

The reason that the search for companions can go deeper than the search
for primaries is that the search space is smaller.  The first task
is therefore to quantify the size of the search space.

I parameterize the semi-major axis $a_c$, and hence the
characteristic angular separation of the companion by
\begin{equation}
\theta_c \equiv {a_c\over a} = \eta\biggl({M\over M_\odot}\biggr)^{1/3}=
\eta\biggl({\rho\over \rho_\odot}\biggr)^{1/3}{D\over D_\odot}\simeq
\eta{D\over D_\odot}
\label{eqn:hill}
\end{equation}
where $D$ and $\rho$ are the primary KBO effective diameter and density
respectively, and where I have approximated $(\rho/\rho_\odot)^{1/3}\sim 1$.
To be bound (within the Hill sphere), $\eta\la 1$.  This sets a strict
upper limit on the separation.  In fact, to remain bound over the
lifetime of the solar system, the companion must be substantially closer,
so this limit is conservative and is therefore robust against somewhat denser
or less reflective KBOs.
Hence, 
$\theta_c\sim \eta \times 1.4^{\prime\prime}(D/10\,\rm km)$ is constrained
to be within a few arcsec near the magnitude 
limit\footnote{When converting from magnitudes to diameters, I adopt
an albedo of 0.04 in R band and a ``typical'' distance of 40 AU, which
yields 11 km at $R=29.6$ corresponding to $H_\vega=28.2$}.  The relative
proper motion of the primary and companion are then of order
\begin{equation}
\Delta\mu\sim \eta^{-3/2}\theta_c \Omega \Pi^{3/2}
\sim {7\,\rm mas\over 72\,\rm day}\eta^{-1/2}{D\over 10\,\rm km}\,
\biggl({a\over 40\,\au}\biggr)^{-3/2}
\label{eqn:hill2}
\end{equation}
That is, near the detection limit (for primaries), one can search
for companions simply by looking for non-moving objects 
(i.e., those that move much less than 1 pixel relative to the
primary during a 72-day campaign)
within a few arcsec of the primary, i.e., $\sim 10^3$ trials for
each primary.  If we estimate that there will be $\sim 10^4$ 
KBO-primary detections,
then $\snr\ga 0.07$ is required to avoid noise-spike contamination.
This implies a flux limit $H_\vega\sim 29.0$, plausibly corresponding
to a diameter $D\sim 7.5\,$km.

For such extremely faint KBO companions, one would measure only their
position.  However, if we consider more generally a pair of (for simplicity)
equal brightness KBOs at a given SNR, their relative proper motion
can be measured to a precision
\begin{equation}
\sigma(\Delta\mu)=\sqrt{24\over N}{\sigma_\ast\over \Delta t}
\rightarrow {35\,\masyr\over\snr}.
\label{eqn:sdmu}
\end{equation}
Then adopting, again for simplicity, \hfil\break
$D=29\,{\rm km}(\snr)^{1/2}$ and
$a=40\,\au$, we expect
\begin{equation}
\Delta\mu\sim 100\,(\snr)^{1/2}\eta^{-1/2}\,\masyr.
\label{eqn:sdmu2}
\end{equation}
This implies, very roughly, that such proper motions can be
detected for $\eta\la (\snr)^3$, i.e., for all bound companions
at $\snr\ga 1$ and for a rapidly declining fraction at fainter magnitudes.

Such proper motion estimates would, by themselves, give crude mass
estimates for individual KBOs.  But the ensemble of such measurements
could be studied statistically to give the mean mass as a function of
(solar system) absolute magnitude.  

Much more detailed orbital motion could be obtained by shallow 
follow-up surveys.  The ensemble of KBOs could be expected to disperse
(relative to mean motion) at roughly $\sim 2\,\kms$, corresponding to
about $0.6^\circ\,\rm yr^{-1}$.  Thus, one could use {\it WFIRST}
itself (in its survey mode) to make brief (e.g., one day) surveys
of the fields to which the KBOs were drifting one to several years
after and/or before the discovery campaign in order to better
characterize the orbits.  The utility and characteristics of
such observations could be much better assessed after logging the
discoveries from the first campaign.

\subsection{{Detection of Unresolved Companions}
\label{sec:unresolved}}

Most KBO binaries detected to date have companions within $\sim 1\,$mag
of the primary, and this does not appear to be the result of selection,
at least for the fairly bright ($R\la 24$) primaries in current 
samples \citep{kbobin}.  Many of these companions are also quite
close, with a median separation near the {\it WFIRST} pixel size $p=110\,$mas.

There are two methods of detecting such close companions that are
of relatively comparably brightness: ``orbiting'' centroid of light
and extended images.  

There are three requirements to detect center of light motion:
First, the actual separation must obey $\theta_c\la p$.  
Otherwise, given the undersampled PSF, the companion would be directly
detectable.  Second, the companion must complete a large fraction
of an orbit.  Otherwise, the centroid of light motion will simply
be absorbed into the KBOs orbital parameters with respect to the Sun.
To be moderately conservative, and for simplicity, I interpret
this requirement as completing one orbit during $\Delta t=72\,$days.
Third, the signal must be sufficiently high to distinguish centroid
motion from noise spikes.

\begin{figure}
\centering
\includegraphics[width=100mm]{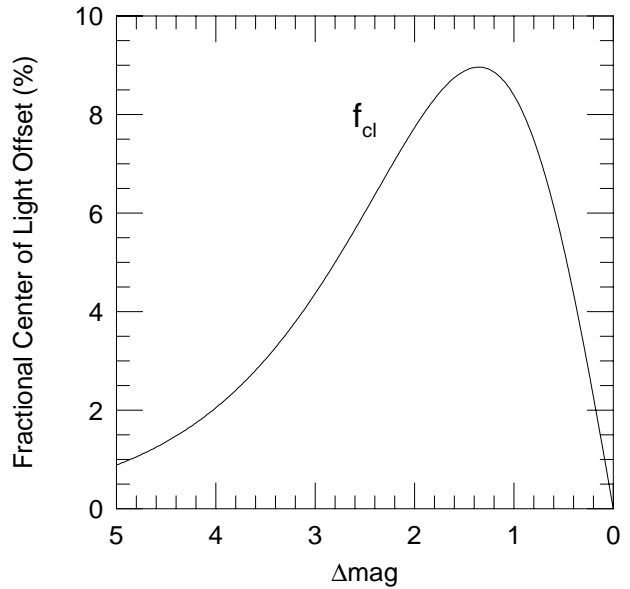}
\caption{Offset of binary center of light from center of mass, as a fraction
of physical separation under the assumptions that the
two components have equal density and equal albedo.  There is
a broad peak $0.7<f_{\rm cl}<0.9$ for $0.72<\Delta{\rm mag}<2.2$.
}
\label{fig:cl}
\end{figure}

Figure~\ref{fig:cl} shows that the fractional displacement of the
centers of light and mass (assuming same density and albedo)
peaks broadly $7\%\la f_{\rm cl}\la 9\%$ for $0.72<\Delta {\rm mag}<2.2$.  
Assuming
that the rms projected amplitude is $\sqrt{2}$ smaller than the
physical separation, and that a $7\,\sigma$ signal is required
for detection, this implies a minimum SNR for each of the 5600
measurements of
\begin{equation}
\snr\ga 7\sqrt{2\over 5600}{p\over f_{\rm cl} \theta_c}\ga 1.65
{p\over \theta_c}\biggl({f_{\rm cl}\over 0.08}\biggr)^{-1},
\label{eqn:cl}
\end{equation}
which corresponds to $H<25.5$  (or $R\la 26.9$).

From the definition of $\eta$,
the boundary $P_c=\Delta t=72\,$days implies 
$\eta=(a/\au)^{-1}(P_c/{\rm yr})^{2/3}=0.0085\,(a/40\,\au)^{-1}$.
Hence, imposing $\theta_c=p$ in Equation~(\ref{eqn:hill}) at this
period yields $D = pD_\odot/\eta = 88\,{\rm km}(a/40\,\au)$,
which corresponds to $H\sim 23.7$ (or $R\sim 25.1$) at $a\sim 40\,\au$.

\begin{figure}
\centering
\includegraphics[width=100mm]{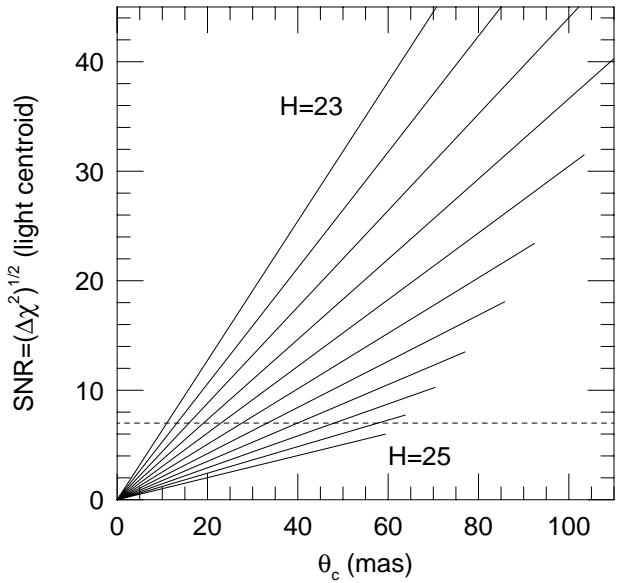}
\caption{Signal-to-noise ratio $[(\Delta\chi^2)^{1/2}]$ for orbiting
KBOs at a range of separations that are less than the WFIRST pixel
size $\theta_c<p=110\,$mas.  The KBO brightness ranges from $H_\vega=23$
to $H_\vega=25$ as indicated.  Tracks end to the right when the orbital
period $P=\Delta t=72\,$days, the duration of an observing campaign.
Longer period orbits would have substantially lower signal.  The
calculations assume $f_{\rm cl}=8\%$ (see Fig.~\ref{fig:cl}) and make
a variety of assumptions listed in the text, including
$D(H_\vega = 24.6)=58\,$km, $a=40\,\au$, and rms projected separation equal to
$\theta_c/\sqrt{2}$.
}
\label{fig:ht}
\end{figure}

These two calculations show that there is considerable parameter space
for detection of this effect.  In Figure~\ref{fig:ht}, I show the
SNR for light centroid motion as a function of binary separation $\theta_c$
for a range of KBO brightnesses from $H=23$ to $H=25$.  The curves
are displayed only for periods $P<\Delta t=72\,$days and so ``cut off''
at the right for the fainter KBO tracks.  Note in particular that
at $H=23$ (roughly $R=24.4$, so almost completely encompassing
the region of present studies), the separation threshold is $\theta_c=11\,$mas,

It is important to remark that this method is critically dependent
on excellent overall {\it WFIRST} astrometry.  For example, the
$\sim 1\,$mas ``light centroid orbits'' for $H=23$ KBOs at the
detection limit in Figure~\ref{fig:ht} would not be detectable
without the excellent world coordinate system described in 
Section~\ref{sec:wcs}.

For truly equal-mass (and equally reflecting) KBOs, the center of mass
and center of light will be identical, implying identically zero
offset between mass and light.  For small differences,
$f_{\rm cl}= 0.11\Delta{\rm mag}$ (see Figure~\ref{fig:cl}).  However,
this near equal-mass regime is the most sensitive to image elongation.
A detailed study of the limits of detectability is beyond the scope
of the present work.  However, I would emphasize that the combined
images for the KBOs in the magnitude range shown in Figure~\ref{fig:ht}
will be fairly deep.  For example, at $H_\vega=25$, the combined image
will have $\snr=200$ and will be resolved at subpixel scales because
it is composed of 5600 dithered images.

\section{{KBO Size Estimates from Occultations}
\label{sec:occult}}

One of the major advantages of KBO searches in dense microlensing
fields is the large number of occultations that are automatically
observed.  These occultations can yield statistical information
on the relation between effective diameter and reflected light,
i.e., the albedo.  In principle, these occultations might give
rise to astrometric noise, but I show that
this effect is negligible.

\subsection{{Occultation Measurements}
\label{sec:signal}}

As I will show below, in contrast to the general astrometric measurements,
which are below sky ($H_\vega<21.7$), the occultations will be
of sources that are above sky.  In particular, these will be substantially
brighter than almost all the KBOs.  Thus, the photon noise will be
dominated by the occulted star.  In addition, there is noise from
unmodeled variability of the KBO.  However, here I will assume that the
variability is well-modeled (at least to the level of the photon
noise for individual measurements) so that only photon noise
need be considered.

However, there is a third form of uncertainty in interpretation of
occultations: the flux of the occulted star.  That is, what
we would like to know is the fraction of the exposure that the
star was occulted, which is just equal to the missing flux divided
by the occulted-star flux.  Hence, for an occultation to be
usefully interpreted, even statistically, the occulted star
must be identified reasonably securely.  The average density of
stars down to $H_\vega<(21,22)$ is $(0.03,0.05)p^{-2}$ (i.e., ``per pixel'').
I assume that sources can be reasonably photometered down to $H_{\vega,*}=21$,
keeping in mind that the ``exact'' location 
of the source (to a small fraction of a pixel) 
is known from the occultation itself together
with the known KBO orbit.  This precise location of the occulted
star (together with subpixel resolution from 40,000 images)
makes it much easier to disentangle the occulted star from significant
blends.

The transverse velocity of the KBO in the Earth frame will generally
be dominated by reflex motion of Earth, which will have an rms value
of $v\sim 10\,\kms$ during the 72 days of observations near
quadrature.  I will adopt this as a typical value in my initial,
simplified treatment.

I begin by examining the case of KBOs with $H=24.6$
(0.5 mag brighter than the break), which
have diameters $D\sim 58\,$km, and hence a maximum self-crossing time
of $D/v\sim 5.8\,$s, i.e., 11\% of an exposure time.  Thus, assuming
that the occultation is entirely contained in the exposure, this
implies a drop of only 11\% of the occulted-star flux.  At $H_*=21$,
the (above-sky) photometric error is $\sim 1.4\%$, implying that
such a flux deficit would be detected at the $11/1.4\sim 8\,\sigma$
level.  This immediately raises the question of what is the
threshold at which we should say that an occultation has been
``detected'' in the face of (negative) noise spikes?  To evaluate
this, we must first estimate the effective number of ``trials''.
I conservatively estimate that $H<21$ star positions can be
measured (independent of the occultation) to 0.3 pixels.  Since
the position of the KBO is known to much higher precision, the
probability that a KBO will land near enough such a star to
be consistent with an occultation is $0.03\times \pi(0.3)^2<1\%$
We must consider all 6900 observations (including $Y$-band and, of course,
those landing near bright stars).  Hence, there will be a total
of $\sim 70$ ``trials'' per KBO.  As I will show shortly, we expect
only about one real occultation from these 70 trials.  This means
that a $3\,\sigma$ cut would yield a $\sim 20\%$ false positive
rate.  This would be acceptable if one carried out a very careful
statistical study but to be conservative, I adopt a $4\,\sigma$ threshold,
for which the (false positive)/(real occultation) rate is about 0.6\%.

Ignoring this threshold for the moment, the rate of occultations
per KBO is
\begin{eqnarray}
& &N_{\rm oc} = {N_{\rm obs}n_{21}Dvt_{\rm exp}\over a^2} 
\nonumber \\
&=&0.6{N_{\rm obs} \over 6900}\,
{n_{21} \over 0.03\,p^{-2}}\,
{D \over 58\,\rm km}\,
{v \over 10\,\kms}\,
{t_{\rm exp} \over 52\,\rm s}\,
\biggl({a \over 40\,\au}\biggr)^{-2},
\label{eqn:occult}
\end{eqnarray}
where $n_{21}$ is the surface density of $H_*<21$ stars.  
In the example that we are considering, the maximum dip is $8\,\sigma$, 
while the adopted threshold $4\,\sigma$.  This means that exposures
that begin more than half-way through an occultation will fall below
the threshold, but these will be exactly compensated by the ones that
end during the occultation but less than half way through it.  More 
generally, such edge effects will not exactly cancel and must be
taken into account.  However, for the present case, Equation~(\ref{eqn:occult})
does not require adjustment.  In addition, KBOs that transit chords
that are less than half the diameter will fall below the detection
threshold.  This corresponds to $1-\sqrt{1 - (4/8)^2}\sim 14\%$ in the
present case (approximating the projected form as a circular disk), but
will vary for other parameters.

In particular, this means that at $H=24.6$, there are about
0.5 occultations per KBO.  As I discuss in
Section~\ref{sec:comp},  this implies that among the several
thousand KBOs discovered, there will be over one thousand occultations.

Because the occultation time is short compared to the exposure
time, the flux decrement (combined with occulted-star flux estimate,
and the known instantaneous KBO proper motion and distance)
directly yields the chord length.  The frequency of such occultations
gives one an estimate of the mean diameter (at fixed absolute magnitude)
while the peak of the chord-length distribution gives another.
Formally, these are
respectively transverse and parallel to the direction of motion,
but these are expected to be statistically identical.  If the typical
detections are $8\,\sigma$, then the second measure will have 8 times
smaller formal error than the first, but the existence of
two independent measurements will provide a useful check on the
systematics.

The KBO break at $H=25.1$ has the peak occultation rate.  At fainter
magnitudes, the KBO luminosity function is flat (and later, perhaps,
falling), while the number of occultations falls as $D$, i.e., as
$10^{-0.2 H}$.  Moreover, at about 1 mag fainter than the break,
the flux decrement drops below the detection threshold (for fiducial
parameters), so one is restricted to brighter (hence rarer) occulted
stars and/or KBOs observed at times that they are moving more slowly
relative to Earth than average.  
On the other hand, KBOs that are 1 mag brighter than
the break are 4 times less common \citep{bernstein04}.  This is
compensated by the factor $\sim 1.6$ larger diameter entering 
Equation~(\ref{eqn:occult}), but still leads to a factor 2.5
(i.e., inversely proportional to flux) fall in the total number
of occultations.

I predict the number of occultations as a function of KBO magnitude
using a more detailed calculation in Section~\ref{sec:comp}.

\subsection{{Occultation Noise}
\label{sec:noise}}

The first point to note is that occulted stars are aligned with the 
geometric center of the KBO much more closely than the astrometric 
precision of individual measurements.  Consider, for example, a
KBO at $40\,\au$ and $H\sim 24.6$ (0.5 mag brighter than 
the break in the luminosity function)
with a diameter of $D\sim 60\,$km.  To be occulted, the star must be
aligned to within $D/2a\sim 1\,$mas, whereas the astrometric precision of
the measurement is only $\sim 25\,$mas.  Hence, the only impact on
measurement is the loss of net flux from the KBO due to a ``hole''
caused by the lost light of the background star.  Moreover, it is only the
{\it absolute value} of the flux change at the KBO position that enters
the astrometric precision.  That is, it is just as easy to centroid
a ``hole'' as a ``bump'', provided it is recognized that there is hole (i.e.,
that an occultation is taking place).  Hence, it is really only
occultations within about a half magnitude of the KBO brightness that
have an adverse impact on astrometry.

To evaluate the practical impact, let us consider two cases, one at $\sim 1$
mag above the break with $H\sim 24$ ($D\sim 75\,$km) and the other near the 
limit $H\sim 28$ ($D\sim 12\,$km).  Since the KBOs are observed near
quadrature, I adopt transverse relative velocities of $v=10\,\kms$,
which implies that the KBOs near the break have a self-crossing time
of about 7 seconds, i.e., $\sim 1/7$ of the exposure times, $t_{\rm exp}=52\,$s.
Hence, for an occultation to reduce the flux by an amount similar to the KBO
brightness requires that the star has $H_*\sim 22$.  The number of 
stars within a half magnitude of this value is similar to the total
number brighter than it, which led (above) to an estimate of less than
one occultation per KBO.

If we now consider KBOs near the limit (i.e., 40 times fainter),
the duration of occultation is $40^{1/2}\sim 6.5$ times shorter,
implying that the ``problematic'' stars are a factor 6.5 (i.e., 2 mag) fainter,
and so only slightly more numerous.  On the other hand, the cross section
is 6.5 times smaller, so the net effect is even smaller.

\section{{Effects of Geosynchronous Orbit}
\label{sec:geo}}

If {\it WFIRST} is in geosynchronous orbit, this will have almost
no effect on the calculations in this paper.  However, it will have
some modest benefits for the followup observations proposed 
in the previous section.

Geosynchronous orbit induces diurnal parallax of amplitude
$\Pi_{\rm geo} = \epsilon\Pi$ where $\epsilon \sim 1/4000$.
Since the target fields lie near the plane of this motion,
diurnal parallax enables a fractional distance measurement
\begin{equation}
\sigma(\Pi)=\sqrt{2\over N}\,{\sigma_\ast\over \epsilon}
\label{eqn:geopi}
\end{equation}
where $N$ is the number of observations in some time that is at
least one day.  Equating this to the second expression in 
Equation~(\ref{eqn:evals1}), one finds that orbital parallax overtakes
geosynchronous parallax at 
$(\Delta t)_\Pi=360^{1/4}\epsilon^{1/2}\Omega^{-1}\sim 4.0\,$day.
A similar calculation for the radial velocity determination (now assuming
at least several days of data) yields a crossover that is just slightly
larger: $(\Delta t)_{v_r} = (35/27)^{1/4}(\Delta t)_\Pi\sim 4.3\,$days.
Thus, diurnal parallaxes are really only useful for the handful of KBOs
that briefly cross the field, or for followup observations that are
carried out over of order one day in non-campaign years.

Similarly, the diurnal parallaxes do not significantly complicate
the search.  Diurnal motion is $\sim \epsilon\Pi\sim 1.3^{\prime\prime}$.
This implies an absolute minimum of about 20 pixels in the parallax
``direction'' of the search space, contrary to the assumption of
the ``position and proper motion'' search described at the beginning of 
Section~\ref{sec:find}.  However, this simplified search was actually
presented only for didactic purposes.  The real searches described
further along in that section already have a much larger parallax 
footprint.  Hence, diurnal parallax due to geosynchronous orbit has
essentially no impact either on the precision of measurement or the
difficulty of searching for KBOs.

\begin{figure}
\centering
\includegraphics[width=100mm]{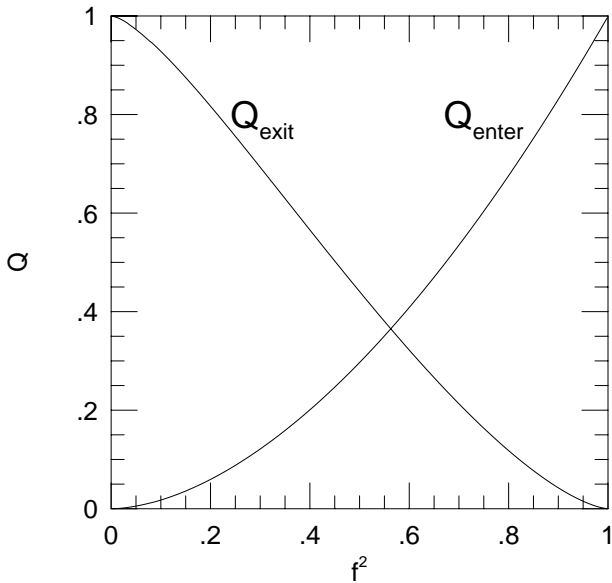}
\caption{Degradation factor $Q$ for the error in the radial velocity
measurement (by far the worst measured phase-space coordinate) as
a function of $f$, the fraction of a campaign that is spent outside 
(inside) the microlensing field by a KBO that exits (enters) during
the campaign.  For low $f$, $Q_{\rm exit}$ is modest for those exiting,
and this is compensated at high $f$ by those entering.  Only about
40\% of all KBOs that are initially in the field exit during a 
campaign, so the net effect of exits/entrances is modest.  Abscissa
is $f^2$ because the distribution of field area is uniform in this
quantity.
}
\label{fig:q}
\end{figure}

\section{{Edge Effects}
\label{sec:edge}}

Up to this point, I have assumed that the KBOs remain in the field for
the entire 72 day campaign.  However, KBOs that are 
initially near the ecliptic East (West) edge of the field for
a Spring (Autumn) campaign will move off the field as Earth approaches
the equinox and will return into the field later on, whereas other KBOs
that initially lie just beyond the ecliptic West (East) edge will
enter the field and then leave it.  Now, as shown in Section~\ref{sec:orbit},
the majority of KBOs do in fact remain in the field for the full 72-day
campaign, and so to zeroth order one could in principle just ignore
these edge effects.  However, here I show that these effects 
actually play only a small role even at first order.

Since the apparent motion of the KBOs is dominated by reflex motion
of Earth, I calculate the time spent within the field as though
this were the only cause.  Then the time spent inside (and outside)
the chip is symmetric about the midpoint of the campaign.  I 
first consider the KBOs that begin within the field and parameterize
the time spent out of the field by $f$, i.e., they spend a time
$f\Delta t$ outside the field.  I focus on the precision of the
radial velocity measurement because it is by far the weakest of
the six phase space coordinates.  Repeating the integral that led to
Equation~(\ref{eqn:cvals}) but excluding observations during the
time spent outside the field, $f\Delta t$, yields
\begin{eqnarray}
&Q_{\rm exit}^2(f)&\equiv{c_{33}({\rm full})\over c_{33}({\rm partial})}
\nonumber \\
&=& {1 - (25/4)f^3(1 - 1.68f^2 + f^4) + f^{10}\over 1 - f^3}.
\label{eqn:qval}
\end{eqnarray}
This function declines monotonically with $f$, but 
there is a compensating effect of KBOs entering the other side of the
field and spending time $f\Delta t$ within the field.  Since these entering
KBOs are observed continuously, the corresponding calculation is trivial,
\begin{equation}
Q_{\rm enter}(f)= f^{3.5} .
\label{eqn:qval2}
\end{equation}
Finally, I note that
one must account for the fact that the distribution
of KBOs leaving (or entering) the field for a time $f\Delta t$ is
not uniform in $f$ but rather in distance from the edge of the field
at the midpoint of the campaign,
which scales $\propto f^2$.  That is, 75\% of all KBOs that leave
(or enter) the field do so for more than half the campaign.

To visualize these effects, I plot $Q_{\rm exit}$ and $Q_{\rm enter}$
versus $f^2$ in Figure~\ref{fig:q}.
When taking account of both effects, the maximum degradation
factor is $Q\sim 0.365$ at $f^2\sim 0.57$.  This factor is modest
given the huge range of KBO brightness being probed.  Moreover,
it is compensated by the fact that twice as many KBOs are probed
at these distances from the edge of the field.  In brief, the
total area of the fields in all six campaigns, 
$6\times 2.8\,{\rm deg}^2 \sim 17{\rm deg}^2$ is a reasonable
estimate of the effective area of the KBO survey.

\section{{KBO Lost and Found}
\label{sec:lost}}

Much of the science that can be extracted {\it WFIRST} KBOs will
be derived directly from the {\it WFIRST} data themselves.  This
includes the distribution of KBOs as a function of orbital parameters,
colors, size, binarity, etc.  However, there are a number of applications
that could require re-observing a detected KBO one or many years later.
For example, as mentioned in Section~\ref{sec:mass}, one might want to
obtain late-time observations of a binary companion in order to
better estimate the corresponding primary's mass.  As another example,
one might want to measure a late-time position of a KBO that has a
moderate-precision period in order to precisely determine whether
it was actually on, or simply near a resonance.

Here I show that all the KBOs discussed above can be unambiguously
recovered even 10 years after they are discovered in {\it WFIRST}
data.  For illustration, I focus on those at the detection limit,
for which I have estimated period errors of $\sigma (P)/P\sim 1.5\%$.
I will then briefly remark on the situation for brighter KBOs.

After 10 years, a KBO at $a\sim 40\,\au$ with 1.5\% period error will
lie somewhere along a ``well-defined arc'', with a $1\,\sigma$
position error along that arc of 
$\sigma(\theta)\sim 1.5\% (10/40^{3/2})\times 360^\circ = 13^\prime$.
Considering that a conservative search might look within $\pm 2.5\,\sigma$,
this implies a search along an arc of $\sim 1^\circ$.  On the other
hand, as discussed below Equation~(\ref{eqn:evals1}), the deviations 
orthogonal to this track are at least 300 times smaller than along it,
implying a search width $\la 12^{\prime\prime}$.  Given that the surface
density of KBOs at this limit is only $\sim 100\,{\rm deg}^{-2}\,{\rm mag}^{-1}$,
there is only a small probability that there will be even one other KBO
of similar magnitude in the search zone.  Because these KBOs are well
below sky, re-detection will necessarily require several epochs and
therefore will automatically return a proper-motion measurement that
will almost certainly distinguish between the (possible) candidates
in this zone.

Brighter KBOs will, of course, be much easier to find.  For example,
at the break in the KBO luminosity function $(H\sim 25.1)$, the period
errors will be smaller by a factor 17, so that the search zone
will be smaller by the same factor in each dimension, i.e., 
$3.5^\prime \times 0.7^{\prime\prime}$.

\section{{Expectations Based on Previous KBO Surveys}
\label{sec:comp}}

\subsection{{Expected KBO Detections}
\label{sec:expfaint}}

The {\it WFIRST} microlensing survey will, without any modification,
probe KBOs down to $H_\vega\sim 28.2$ over an area of $\sim 17\,{\rm deg}^2$
and yield orbits with period precision of 1.5\% at the magnitude
limit (and much better at brighter magnitudes).  For my adopted
conversion $R-H=1.4$, 
this corresponds to $R\sim 29.6$.  How does this compare to
previous deep surveys?

\begin{figure}
\centering
\includegraphics[width=100mm]{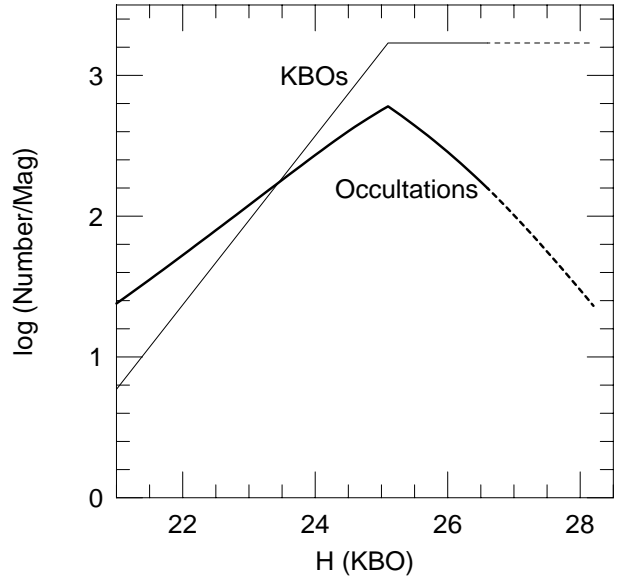}
\caption{Predicted number of KBO detections (solid) and KBO
occultations (bold) per magnitude that can be extracted from
the {\it WFIRST} microlensing survey.  The detections are
derived by multiplying the \citet{bernstein04} KBO luminosity
function by $17\,{\rm deg}^2$.  The curve is dashed for $H>27.1$
because the KBO frequency is not presently known in this regime.
The calculation of the occultation rate per KBO is outlined in
Sections~\ref{sec:signal} and \ref{sec:expoccult}.  In particular,
I count only occultations of $H_*<21$ stars that can be detected
with at least $4\,\sigma$ significance.
}
\label{fig:rate}
\end{figure}

\citet{bernstein04} searched $0.019\,{\rm deg}^2$ down to $R\leq 29$
(strictly, $m_{606}\leq 29.2$) using 22 ks exposures with ACS on the 
{\it HST}.  They found three new objects.  For
the two of these with $m_{606}>28$, they obtained only crude ($\sim 30\%$)
orbital parameters.  They found no KBOs $m_{606}>28.3$ even though
they had near-100\% completeness to $m_{606}=29$, implying that
the distribution is flat (or falling) beyond $R=28.5$.  
Combining their results with previous work at brighter magnitudes,
they fit a relation that can be expressed as,
\begin{equation}
{d^2N\over dR\,d\Omega} = {100\over \rm mag\,deg^2}\,10^{0.6(R-26.5}
\qquad (R\leq 26.5)
\label{eqn:bern}
\end{equation}
and then a roughly flat distribution 
$\sim 100\,{\rm mag}^{-1}{\rm deg}^{-2}$ for $26.5<R<28.5$ 
(and possibly beyond).  
The rising part of this distribution contains $\sim 70\,{\rm deg}^{-2}$.
Based on this estimate, we can expect that {\it WFIRST} would detect
4500--6500 KBOs.  The \citet{bernstein04} relation (Equation~(\ref{eqn:bern}))
scaled by $17\,{\rm deg}^2$ is shown in Figure~\ref{fig:rate}.  The
dashed part of the curve indicates an extension into the unmeasured regime.

\subsection{{Expected Occultations}
\label{sec:expoccult}}

In Section~\ref{sec:occult}, I presented a simplified calculation
of the KBO occultation rate, in order to illustrate the
basic physics.  In Figure~\ref{fig:rate}, I show the results of
a more detailed calculation including all the elements outlined
in Section~\ref{sec:occult}.  In particular, I integrate over the
full range of Earth-KBO relative velocities, which is mainly
driven by the changing reflex motion of Earth but also includes
a small component due to intrinsic KBO motion.  I continue to demand
$4\,\sigma$ detections and consider only occultations of sources $H\leq 21$.
The principal results from the simplified calculations are all confirmed.

One point to further note is that at the bright end of the distribution
shown in Figure~\ref{fig:rate} ($H=21$) there are about four occultations
per KBO.  Although this applies to only a half dozen objects per magnitude,
it does hold out the hope that some shape information can be extracted,
particularly if high-resolution followup imaging can determine the
precise source location and therefore the precise impact parameter of
the occultation.

\subsection{{Expected Binary KBO Detections}
\label{sec:expbin}}

Statistics on KBO companions are available primarily for relatively
bright primaries $R\la 24$, corresponding to $H\la 22.6$.  These have
a median separation of $\sim 100\,$mas, and are primarily of near-equal
brightness, with a substantial majority roughly uniformly distributed
over $1>\Delta{\rm mag}>0$.  About 22\% of classical and 5\% of other
KBOs have such companions \citep{kbobin}.
Essentially all analogs of these companions will be found
by {\it WFIRST} for the bright KBOs in its field.  If there are similar
companions down to $H=24$, then the total number of such binaries
will be about 200.  As shown by Figure~\ref{fig:ht}, at fainter
magnitudes the close binaries become less accessible and then 
inaccessible for $H>25$.  Moreover, there is essentially no information
on the frequency of companions at these magnitudes.  Hence, while
the unexplored parameter space is fairly large, there is no
reliable way to estimate the number of companions in these regimes.

\section{{Application to Ground-Based Microlensing Surveys}
\label{sec:ground}}


There are several ground-based microlensing surveys that could in
principle be searched for KBOs, including the ongoing OGLE-IV and MOA-II
surveys, as well as the KMTNet survey, which is about to begin.
Here I describe some relatively low-effort ``entry points'' into
these data sets and briefly sketch extensions that would probe
much deeper.  The ``entry point'' searches could be carried out
using a single night of data and would yield of order a dozen KBOs.  The
extensions could reach within $\sim 2\,$mag of the break in the 
luminosity function $R\sim 26.5$, thus multiplying this number several fold.

The calculations below are based on the summaries provided by
\citet{henderson14} of the telescope, detector, and observing characteristics
of the OGLE and KMTNet surveys.  In particular
I adopt a photo-electron rate $\dot\gamma=4.91(D/1.6{\rm m})^2 {\rm s}^{-1}$
at $I=22$, where $D$ is the diameter of mirror, an ambient background
of $I=18.8\,{\rm mag\,arcsec}^{-2}$, and an effective PSF area of
$\Omega_{\rm back} = 1.7\times 4\pi{\rm (FWHM)}^2/\ln 256$.  I also assume
that the below-sky errors are 1.3 times larger than the photon noise.
As noted by \citet{henderson14}, the $I=18.8$ background, based on current
OGLE-IV data, is not understood, and might be improved by further
technical developments.  Hence, these numbers are conservative.

OGLE observes three fields, totaling about $5\,{\rm deg}^2$, with
a cadence of $3\,{\rm hr}^{-1}$, with exposures of about 100 s, using
a 1.3m telescope.
The fields can be observed for 10 hours per night (so 30 observations)
for about one month centered on the summer solstice.  I assume that
one of these nights is clear, with low moonlight, and very good seeing
of FWHM$\sim 1^{\prime\prime}$, and that 90\% of the observations are not
seriously affected by above-sky stars, cosmic rays, or pixel defects.
Each (below-sky) observation then has $\snr = 10^{0.4(22.5-I)}$. The 
observations take place at opposition rather than quadrature (as for
{\it WFIRST}), which means one must consider a proper motion rectangle
of $(30\,\kms/30\,\au)\times(3\,\kms/30\,\au)$, which for 10
hours of observations and $1^{\prime\prime}$ seeing implies 300 searches
per resolution element or a total of $2\times 10^{11}$ searches.
This requires a total $\snr=7$, which in $30\times 0.9=27$ observations
can be achieved for $I=22.2$ KBOs, corresponding to
$R=22.7$.  There are roughly 0.4 such KBOs per square degree, implying
that a few would be expected in the $5\,{\rm deg}^2$ high-cadence OGLE
field.  However, this could be repeated for each of 5 years of OGLE-IV
archival data, yielding about 10 KBOs.

KMTNet will have three 1.6m telescopes, two of which will be at 
excellent and somewhat overlapping sites in Chile and South Africa.
Its cadence at each will be $6\,{\rm hr}^{-1}$, with
120 s exposures over $16\,{\rm deg}^2$.  Combining
these and adopting a slightly worse mean seeing of $1.2^{\prime\prime}$
(to account for the greater difficulty of coordinating observations
from two sites), yields a similar $\snr = 10^{0.4(22.6-I)}$. Then, taking
account of the four times greater number of exposures, the limit
of detectability is $I\sim 23.0$ corresponding to $R=23.5$.  This
would yield roughly 1 KBOs per square degree or a total of about 15.

Once identified, these KBOs would yield excellent orbits because
microlensing fields cover about $100\,{\rm deg}^2$ over many years,
albeit at lower cadence.

Over a whole season, KMTNet would observe any given KBO roughly 10,000
times from Chile and South Africa, of which 1/4 would be in good seeing,
good transparency, and with low Moon background.  As with {\it WFIRST}, the
search space would be much larger than for a simple one-night search, so
I adopt a similar $\Delta\chi^2=115$ threshold.  This leads to a detectability
limit of $I=24.3$, or $R\sim 24.8$, i.e., about 100 KBOs.  Note that in
contrast to {\it WFIRST} detections, one need not be extremely rigorous
about eliminating noise spikes at the detection phase because these
will be automatically vetted when the KBO is tracked to additional
seasons to measure its orbit.

\section{{Conclusions}
\label{sec:conclude}}

Space-based microlensing surveys are an extremely powerful probe
of KBOs basically because microlensing motivates very high cadence
observations over long time baselines and fairly wide fields that
happen by chance to lie near the ecliptic.  The very large number
of images allows one to construct essentially noiseless (compared
to the individual images) templates, and so construct essentially
blank images from the ``crowded'' fields via image subtraction.

In particular, the {\it WFIRST} microlensing survey, without any
modification, can yield 4500-6500 KBOs down to $H_\vega=28.2$.
The last magnitude of such a search requires algorithmic and/or
hardware development to carry out the computations in a timely
manner.  However, the more restricted search to $H_\vega=27.1$
(with 4500-5000 KBOs) can be carried out by simple brute force
searches using today's technology.

Because the detections arise from a near-continuous time series
over much less than a year, and centered at quadrature, the
statistical characterization of the orbital parameters is best
understood in a 6-D cartesian framework.  The same framework allows
one to directly map the expected (or, more generally, allowed)
orbit space into a cartesian search space.  In particular, I find
that for {\it WFIRST}, the period errors scale as 
$\sigma(P)/P\sim 0.09\%\times 10^{0.4(H_\vega-H_{\rm break})}$, where
$H_{\rm break}=25.1$ is the break in the luminosity function.

Binary companions that are separated by a few pixels can be found down to
about $H_\vega\sim 29$, regardless of the limit of the primary search.
The limit is deeper because the search space is smaller, implying
fewer noise spikes.  These binaries can provide statistical mass information,
or if followed up by additional observations, individual mass
measurements.

Binary companions with separations down to 0.1 pixels (11 mas) can be found
for roughly equal (but not exactly equal) masses for primaries
$H\leq 23$ from the offset between centers of mass and light,
and for larger subpixel separations down to $H\sim 25$.  Exactly
(or very nearly) equal-mass binaries at sub-pixel separations
can be detected from image elongation.  Analogs to essentially
all binaries currently being found ($R\la 24$, $\theta_c\ga 0.01^{\prime\prime}$)
will be found by {\it WFIRST}, but it will also probe a huge parameter
space of binary companions that has not yet been explored.

A side benefit of the fact that microlensing searches are carried
out in the most crowded fields (prior to image subtraction)
is the high probability of occultations.
On average, each KBO at the break will occult 0.4 stars with $H_*<21$
(so reliably detected in the deep drizzled image) and with at least
$4\,\sigma$ detections.  Over 1000 occultations of detected KBOs will
enable measurement of the KBO albedo as functions of orbital
properties and absolute magnitude.

Finally, using the same techniques outlined in this paper, it should
be possible to find roughly 100 KBOs using current and soon-to-be-initiated
ground-based microlensing surveys.



\acknowledgments

I thank Matthew Penny, Scott Gaudi, and Radek Poleski for helpful suggestions.
This work was supported by NSF grant AST 1103471 and NASA grant NNX12AB99G.




\begin{thebibliography}{}


\bibitem[Alard \& Lupton(1998)]{alard98} Alard, C. \& Lupton, R.H.,
1998, \apj, 503, 325

\bibitem[Bernstein et al.(2004)]{bernstein04}
Bernstein, G.M., Trilling, D.E., Allen, R.L.\ 2004, \aj, 128, 1364

\bibitem[Doressoundiram et al.(2008)]{kbocolor}
Doressoundiram, A., Boehnhardt, H., Tegler, S. C., \& Trujillo, C.\ 2008,
in The Solar System Beyond Neptune, M.A. Barucci, H. Boehnhardt, D.P. 
Cruikshank, and A. Morbidelli (eds.), University of Arizona Press, 
Tucson, p.91-104

\bibitem[Doressoundiram et al.(2008)]{kbobin}
Noll, K.S., Grundy, W.M., Chiang, E.I., et al.\ 2008,
in The Solar System Beyond Neptune, M.A. Barucci, H. Boehnhardt, D.P. 
Cruikshank, and A. Morbidelli (eds.), University of Arizona Press, 
Tucson, p.345-363

\bibitem[Gomes et al.(2005)]{nice} Gomes, R., Levison, H.F., Tsiganis, K.,
\& Morbidelli, A. 2005, Natur, 435, 7041

\bibitem[Gould(1996)]{gould96} Gould, A.\ 1996, \apj, 470, 201

\bibitem[Gould(2004)]{gould04} Gould, A.\ 2004, arXiv:astro-ph/0403506

\bibitem[Gould \& Yee(2013)]{gould13} Gould, A.\ \& Yee, J.C.\ 2013,
\apj, 767, 42

\bibitem[Henderson et al.(2014)]{henderson14} Henderson, C.B., 
Gaudi, B.S., Han, C., et al.\ 2014, \apj, submitted

\bibitem[Monet et al.(2003)]{usnob} Monet, D.G., Levine, S.E., Canzian, B.,
et al.\ 2003, \aj, 125, 984


\bibitem[Slipher(1930a)]{slipher30} Slipher, V.M.\ 1930, Lowell Observatory 
Observation circular (March 13)

\bibitem[Slipher(1930b)]{slipher30b} Slipher, V.M.\ 1930b, Popular Astronomy,
38, 187

\bibitem[Shao et al.(2014)]{shao14} Shao, M., Nemati, B., Zhai, C.\ 2014,
\apj, 782, 1

\bibitem[Shephard et al.(2011)]{ogleplane} 
Shephard, S.S., Udalski, A., Trujillo, C., et al.\ 2011, \aj, 142, 98



\end{thebibliography}
\end{document}